\documentstyle{article}
\begin{document}  
\title{Forward-Backward Asymmetry in $B\rightarrow\  X_d e^+e^-$ }
\author{ S.Rai Choudhury\\ I.C.T.P., I-34100 TRIESTE \\ and\\Department
of Physics, University of Delhi,Delhi 110007, India.} 
\maketitle
                                   
\begin{abstract}
The Forward-backward asymmetry in the angular distribution of $e^+e^-$ is
studied in the processes $B\rightarrow\ X_d e^+ e^-$ and
$\bar{B}\rightarrow\ \bar{X}_d e^+ e^-$. The possibility of observing CP-
violation through the asymmetries in these two processes is examined.
\end{abstract}

\newpage

    The flavor changing neutral current transitions  $B\rightarrow\ X_s e^+ e^-$ and
    $B\rightarrow\ X_d e^+ e^-$  offer a deeper proble  for the weak interaction
    sector of the standard model since they go through second  order weak
interactions.  The basic quark transition involved in these  $b\rightarrow\ s$
and  $b\rightarrow\  d$ occur through an intermediate t,\ c,\ or u\ quark.
    These processes can be described in terms of an
     effective Hamiltonian which     incorporates both the results of  short
distance expansion techniques as well as the
effect
    of virtual quark-antiquark pairs \cite{Buch} . For $B\rightarrow\ X_s
e^+
e^-$ the 
    transitions involving intermediate
    top, charm  and u-quarks\  enter respectively with factors\  $V_{tb}
V^*_{ts},\
    V_{c b}V^*_{cs}\   and\  V_{ub} V^*_{us}$ . The last of these three is
extremely
small    compared to the other two ; by the unitarity relation between the elements of
the CKM
    matrix, the first two of these become effecively  negative of each other. Thus
the CKM factors
    effectively act as an overall factor with the result that there is very little
    chance of obtaining details of the CKM matrix, in particular its CP-violating phase
    from the study of this process. Kr\"{u}ger and Sehgal \cite{Krug} have
recently pointed out
that
    the situation is quite different for the process $B\rightarrow\ X_s e^+ e^-$.\ 
    There the three CKM factors,   obtained from above by the
    replacement s$\rightarrow$d , are comparable so that the cross section for
    the
    process will have significant interference terms, possibly opening up prospects
    of meaningful estimation of the complex CKM matrix elements.                                                                                      
 Kr\"{u}ger and Sehgal \cite{Krug} have shown that the cross-section for
the
processes
$B\rightarrow\
    X_d e^+ e^-  and\   \bar {B} \rightarrow\ \bar{X}_d e^+ e^-$ have differences
    depending
on    the value of the CKM matrix elements that may be experimentally significant
in the near future.

For inclusive B-decays into lepton pairs, there is another asymmetry, namely the
forward-backward (FB) asymmetry, introduced by Ali,Mannel and
Morozumi \cite{Ali},which
again is another parameter which is likely to be very useful for comparison
of theory with experimental data. For $B\rightarrow\ X_se^+e^-$ \  process, this
parameter
is again not very sensitive to the CKM matrix elements and furthermore tha
magnitude of this parameter is the same for the $\bar{B}$ \ decay as for the
B-decay. The observations made by Kr\"{u}ger and Sehgal[2] however show
that we may
expect that these parameters will be more sensitive in the inclusive B-decay
into lepton pair with non-strange hadrons. 
 In this brief note we obtain quantitative predictions for the FB asymmetry for
the processes $B\rightarrow\ X_d\ e^+\ e^-$ \  and $\bar{B}\rightarrow\  \bar{X_d}
\ e^+\ e^-$.\  The results as expected show considerably more dependence on the
value of the CKM matrix elements and also brings out the possibility of observing
the CP-violating phase of the CKM matrix elements in a mixture of equal numbers
of $B$ and $\bar{B}$ particles.

In the lowest order of the heavy quark effective theory, the process \ $B\rightarrow\ X
_d\ e^+\ e^-$ can be equated to the QCD corrected matrix element for the process ( with
the p's\   and \ q's\ representing the on-shell momentum of the particles)

\centerline{$b\ (p\ _b)\ \rightarrow\ d\ 
(p_d)\  +\ e^+\ (q\ _1\ )\ +\ e^-\ (\ q\ _2)\ $}

The standard kinematical variables for this process are ( with all dimensional
quantities scaled to the b-quark mass):

\centerline{$q=\ q_1\ +\ q_2\ $\ ;\ $s\ =(\ q^2\ )$\ ;\ $u=2\ p_b.\ (q_2\
-q_1)$}
\centerline{$u^2(\ s\ )\ =\ (s\ -\ \left(1\ +m_d^2\ )\right)\ (\ s\ -\ \left
(\ 1\ -m_d^2\ )\right)\ ;\  z\ =\ u\ /\ u(\ s\ )$}
z is the cosine of the angle between $\vec{q_1}$ \ and \ $\vec{p_b}$ in the rest frame
of the lepton pair. In terms of these variables and the standard Wilson coefficients
$C_i$\ 's,[1] the matrix element for the above process can be written as:
\begin{equation} 
M\ (p_b,q_1,q_2\ ;\ \lambda\ )=2\ 
\sqrt{2}\ G_F\ V_{tb}\ V^*_{td}\ \ (\alpha\ /4\pi)\ F,
\end{equation}
where
\begin{eqnarray} 
F  & =  &C_9^{eff}\ (\bar{d}\ \gamma_\mu\ b_L)\ (\bar{e}\ \gamma_\mu\ e)
              + C_{10}\ (\bar{d}\ \gamma_\mu\ b_L)\ (\bar{e}\
\gamma^\mu\ \gamma^5\ e)\nonumber \\
   &    & -\ 2\ C_7\ [\bar{d}\ i\ \sigma_{\mu\nu}]
         {(\ m_b\ b_R\ +\ m_d\ b_L\ )}\ (\ \bar{e}\ \gamma^\mu\ e\ ))(q^\nu\ /q^2)
\end{eqnarray}
The constant C's  are given by\ [1]

\centerline{        $ C_1= -0.249,$\          $C_2= 1.108,$\       $C_3=
1.112 X 10^{-2}$} 

\centerline{   $C_4=-2.569 X10^{-2}$\  , $C_5= 7.404 X 10^{-3}$,\
$C^6= -3.144 X10
^{-2}$}
\centerline{  $C_7 = -0.315$,\      $C_9 = 4.227 ;$}

$C_9^{eff}$ is given by:
\begin{equation}
C_9^{eff}\ =\ \xi_1\ +\lambda\ \xi_2,\
\end{equation}
 with

\begin{eqnarray}
\xi_1 & =& C_9
+g_c\ (\ 3C_1\ +C_2\ +3C_3\ +C_4+\  3C_5\ C_6\ )\nonumber \\
      &  &
 -\ 1/2\ g_d\ (C_3\ +\ 3C_4)\ -\ 1/2\ g_b\ (4C_3+\ 4C_4\ +3C_5+\ \ C_6)\nonumber \\
      &  &
\ +2\ (3C_3+\ C_4+\ 3C_5+\ C_6)/9
\end{eqnarray}
\[ \xi_2\ =\ (g_c\ -g_u\ )\ (3\ C_1\ +\ C_2\ )\]  

where

\begin{eqnarray}
g_q &=& -8\ ln(m_q) /9\ +8/27\ +4\ y_q/9\ -(4+2y_q)\ /3+\nonumber \\
    & & +\sqrt{x}\ \theta(x)\left[\ ln(1+x)\ -ln(1-x)\
-i\pi\right]\nonumber\\
    & & +\sqrt{-x}\theta(-x)[\ 2\   arctan (1/\sqrt{-x})],
\end{eqnarray}
with \ $y_q\ =4\ m_q^2\ /s\ ;\ x=1-\ y_q$.\ 
The parameter $\lambda$ \ is the ratio $V_{ub}V^*_{ud}\ /V_{tb}V^*_{td}$ and can
be expressed in terms of the Wolfenstein parameters as:
\begin{equation}
\lambda=\ \frac{\rho(1-\rho)-\eta^2\ -i\ \eta}{\ (1-\rho)^2\ +\eta^2}
\end{equation}
Large distance effects can also be included in this scheme by adding to the
expression for  $C_9^{eff}$ \ suitable
Breit-Wigner forms corresponding to the $J/\psi,\ \psi'$ resonances. However
as we will see, the region in which we will be interested in is below the
resonance region and we will therefore neglect them.
We shall also set the mass of the d-quark  and the electron zero. 
 With the matrix elemtent
given as above, the differential cross-section for the process $b\rightarrow\
d+e^+e^-$  \  can be worked
out as 
\begin{eqnarray}
\frac{d^2\sigma}{dz\ ds}& = & C\ (1\ -\ s\ )\ \big(\ (|C_9^{eff}|^2\ +\
C_{10}^2\ )\ \ [\ 2\ +
                              2s+\ -\ 2 z^2(1-s)\ ] \nonumber\\
                      & &   + 8\ C_7^2\ [2\ -(\ 1\ -s\ )\ (1-\
z^2)\ ]\ /s\nonumber\\
                      & &   +8\  Re(C_9^{eff})\ [2\ C_7\ -s\ z\ C_{10}\ ]
\nonumber\\
                      & &      -16\ C_{10}\ C_7\ z \big)
\end{eqnarray}
where C is an overall constant.
\newline
With this expression the normalized FB asymmetry \ A(s)\ defines as
\begin{eqnarray}
A(s)& = &     \frac{[\int_{0}^{1}\ -\int_{-1}^{0}]\ dz\ D(z,s)}
                 {[\int_{0}^{1}\ +\int_{-1}^{0}]\ dz\ D(z,s)}\nonumber\\
    & = & \frac{-3[\ s\ C_{10}\ reC_9^{eff}\ +2 C_{10}\ C_7\ ]}
               {(1+2s)\ [\ |C_9^{eff}|^2\ +C_{10}^2\ ]\ +4C_7^2\ (2+1/s)\
+
                12\ C_7\ reC_9^{eff}}    
\end{eqnarray}
where $D(z.s)$ is the left hand side of equation ( 6 ). 

The above expression refers to the transion $b(p_b)\rightarrow\ d\
e^+(q_1)e^-(q_2)$. By the
CPT theorem, the matrix element for the process
$\bar{b}(p_b)\rightarrow\ \bar{d}\  e^-(q_1)e^+(q_2)$ 
is given by $M(\ p_b,\ q_2,\ q_1\ ;\ \lambda^*\ )$.\ Thus but for the imaginary part
part of $\lambda$\ ,equation ( 6 ), the FB asymmetry of the process
$\bar{B}\rightarrow\
\bar{X}_d\ 
e^+\ e^-\ $ would be exactly the negative of B - decay.  The difference in the 
magnitude of the B- and the $\bar{B}$ FB asymmetry would thus directly measure the
CP violating phase of the CKM matrix. 

Figures (1)-(3) show the calculated values of the two asymmetry parameters
\begin{figure}[tbh]
\begin{center}
\caption{FB asymmetry for $\rho$=0.30}
% GNUPLOT: LaTeX picture
\setlength{\unitlength}{0.240900pt}
\ifx\plotpoint\undefined\newsavebox{\plotpoint}\fi
\sbox{\plotpoint}{\rule[-0.200pt]{0.400pt}{0.400pt}}%
\begin{picture}(750,720)(0,0)
\font\gnuplot=cmr10 at 10pt
\gnuplot
\sbox{\plotpoint}{\rule[-0.200pt]{0.400pt}{0.400pt}}%
\put(176.0,383.0){\rule[-0.200pt]{122.859pt}{0.400pt}}
\put(176.0,68.0){\rule[-0.200pt]{0.400pt}{151.526pt}}
\put(176.0,68.0){\rule[-0.200pt]{4.818pt}{0.400pt}}
\put(154,68){\makebox(0,0)[r]{-0.3}}
\put(666.0,68.0){\rule[-0.200pt]{4.818pt}{0.400pt}}
\put(176.0,173.0){\rule[-0.200pt]{4.818pt}{0.400pt}}
\put(154,173){\makebox(0,0)[r]{-0.2}}
\put(666.0,173.0){\rule[-0.200pt]{4.818pt}{0.400pt}}
\put(176.0,278.0){\rule[-0.200pt]{4.818pt}{0.400pt}}
\put(154,278){\makebox(0,0)[r]{-0.1}}
\put(666.0,278.0){\rule[-0.200pt]{4.818pt}{0.400pt}}
\put(176.0,383.0){\rule[-0.200pt]{4.818pt}{0.400pt}}
\put(154,383){\makebox(0,0)[r]{0}}
\put(666.0,383.0){\rule[-0.200pt]{4.818pt}{0.400pt}}
\put(176.0,487.0){\rule[-0.200pt]{4.818pt}{0.400pt}}
\put(154,487){\makebox(0,0)[r]{0.1}}
\put(666.0,487.0){\rule[-0.200pt]{4.818pt}{0.400pt}}
\put(176.0,592.0){\rule[-0.200pt]{4.818pt}{0.400pt}}
\put(154,592){\makebox(0,0)[r]{0.2}}
\put(666.0,592.0){\rule[-0.200pt]{4.818pt}{0.400pt}}
\put(176.0,697.0){\rule[-0.200pt]{4.818pt}{0.400pt}}
\put(154,697){\makebox(0,0)[r]{0.3}}
\put(666.0,697.0){\rule[-0.200pt]{4.818pt}{0.400pt}}
\put(176.0,68.0){\rule[-0.200pt]{0.400pt}{4.818pt}}
\put(176,23){\makebox(0,0){0}}
\put(176.0,677.0){\rule[-0.200pt]{0.400pt}{4.818pt}}
\put(249.0,68.0){\rule[-0.200pt]{0.400pt}{4.818pt}}
\put(249,23){\makebox(0,0){0.05}}
\put(249.0,677.0){\rule[-0.200pt]{0.400pt}{4.818pt}}
\put(322.0,68.0){\rule[-0.200pt]{0.400pt}{4.818pt}}
\put(322,23){\makebox(0,0){0.1}}
\put(322.0,677.0){\rule[-0.200pt]{0.400pt}{4.818pt}}
\put(395.0,68.0){\rule[-0.200pt]{0.400pt}{4.818pt}}
\put(395,23){\makebox(0,0){0.15}}
\put(395.0,677.0){\rule[-0.200pt]{0.400pt}{4.818pt}}
\put(467.0,68.0){\rule[-0.200pt]{0.400pt}{4.818pt}}
\put(467,23){\makebox(0,0){0.2}}
\put(467.0,677.0){\rule[-0.200pt]{0.400pt}{4.818pt}}
\put(540.0,68.0){\rule[-0.200pt]{0.400pt}{4.818pt}}
\put(540,23){\makebox(0,0){0.25}}
\put(540.0,677.0){\rule[-0.200pt]{0.400pt}{4.818pt}}
\put(613.0,68.0){\rule[-0.200pt]{0.400pt}{4.818pt}}
\put(613,23){\makebox(0,0){0.3}}
\put(613.0,677.0){\rule[-0.200pt]{0.400pt}{4.818pt}}
\put(686.0,68.0){\rule[-0.200pt]{0.400pt}{4.818pt}}
\put(686,23){\makebox(0,0){0.35}}
\put(686.0,677.0){\rule[-0.200pt]{0.400pt}{4.818pt}}
\put(176.0,68.0){\rule[-0.200pt]{122.859pt}{0.400pt}}
\put(686.0,68.0){\rule[-0.200pt]{0.400pt}{151.526pt}}
\put(176.0,697.0){\rule[-0.200pt]{122.859pt}{0.400pt}}
\put(176.0,68.0){\rule[-0.200pt]{0.400pt}{151.526pt}}
\put(556,632){\makebox(0,0)[r]{For B}}
\put(578.0,632.0){\rule[-0.200pt]{15.899pt}{0.400pt}}
\put(249,251){\usebox{\plotpoint}}
\multiput(249.00,251.58)(0.594,0.498){71}{\rule{0.576pt}{0.120pt}}
\multiput(249.00,250.17)(42.805,37.000){2}{\rule{0.288pt}{0.400pt}}
\multiput(293.58,288.00)(0.498,0.569){83}{\rule{0.120pt}{0.556pt}}
\multiput(292.17,288.00)(43.000,47.846){2}{\rule{0.400pt}{0.278pt}}
\multiput(336.58,337.00)(0.498,0.568){85}{\rule{0.120pt}{0.555pt}}
\multiput(335.17,337.00)(44.000,48.849){2}{\rule{0.400pt}{0.277pt}}
\multiput(380.58,387.00)(0.498,0.545){85}{\rule{0.120pt}{0.536pt}}
\multiput(379.17,387.00)(44.000,46.887){2}{\rule{0.400pt}{0.268pt}}
\multiput(424.58,435.00)(0.498,0.511){83}{\rule{0.120pt}{0.509pt}}
\multiput(423.17,435.00)(43.000,42.943){2}{\rule{0.400pt}{0.255pt}}
\multiput(467.00,479.58)(0.536,0.498){79}{\rule{0.529pt}{0.120pt}}
\multiput(467.00,478.17)(42.901,41.000){2}{\rule{0.265pt}{0.400pt}}
\multiput(511.00,520.58)(0.579,0.498){73}{\rule{0.563pt}{0.120pt}}
\multiput(511.00,519.17)(42.831,38.000){2}{\rule{0.282pt}{0.400pt}}
\multiput(555.00,558.58)(0.629,0.498){67}{\rule{0.603pt}{0.120pt}}
\multiput(555.00,557.17)(42.749,35.000){2}{\rule{0.301pt}{0.400pt}}
\multiput(599.00,593.58)(0.652,0.497){63}{\rule{0.621pt}{0.120pt}}
\multiput(599.00,592.17)(41.711,33.000){2}{\rule{0.311pt}{0.400pt}}
\multiput(642.00,626.58)(0.760,0.497){55}{\rule{0.707pt}{0.120pt}}
\multiput(642.00,625.17)(42.533,29.000){2}{\rule{0.353pt}{0.400pt}}
\put(556,587){\makebox(0,0)[r]{For $\bar{B}$}}
\multiput(578,587)(20.756,0.000){4}{\usebox{\plotpoint}}
\put(644,587){\usebox{\plotpoint}}
\put(249,496){\usebox{\plotpoint}}
\multiput(249,496)(15.014,-14.331){3}{\usebox{\plotpoint}}
\multiput(293,454)(13.533,-15.737){4}{\usebox{\plotpoint}}
\multiput(336,404)(13.712,-15.581){3}{\usebox{\plotpoint}}
\multiput(380,354)(14.347,-14.999){3}{\usebox{\plotpoint}}
\multiput(424,308)(14.848,-14.503){3}{\usebox{\plotpoint}}
\multiput(467,266)(15.532,-13.767){3}{\usebox{\plotpoint}}
\multiput(511,227)(16.243,-12.921){2}{\usebox{\plotpoint}}
\multiput(555,192)(16.786,-12.208){3}{\usebox{\plotpoint}}
\multiput(599,160)(17.393,-11.326){2}{\usebox{\plotpoint}}
\multiput(642,132)(18.731,-8.940){3}{\usebox{\plotpoint}}
\put(686,111){\usebox{\plotpoint}}
\sbox{\plotpoint}{\rule[-0.400pt]{0.800pt}{0.800pt}}%
\put(556,542){\makebox(0,0)[r]{For B+$\bar{B}$}}
\put(578.0,542.0){\rule[-0.400pt]{15.899pt}{0.800pt}}
\put(249,378){\usebox{\plotpoint}}
\put(249,374.84){\rule{10.600pt}{0.800pt}}
\multiput(249.00,376.34)(22.000,-3.000){2}{\rule{5.300pt}{0.800pt}}
\put(293,371.84){\rule{10.359pt}{0.800pt}}
\multiput(293.00,373.34)(21.500,-3.000){2}{\rule{5.179pt}{0.800pt}}
\put(336,369.34){\rule{10.600pt}{0.800pt}}
\multiput(336.00,370.34)(22.000,-2.000){2}{\rule{5.300pt}{0.800pt}}
\put(380,367.34){\rule{10.600pt}{0.800pt}}
\multiput(380.00,368.34)(22.000,-2.000){2}{\rule{5.300pt}{0.800pt}}
\put(424,365.34){\rule{10.359pt}{0.800pt}}
\multiput(424.00,366.34)(21.500,-2.000){2}{\rule{5.179pt}{0.800pt}}
\put(467,363.84){\rule{10.600pt}{0.800pt}}
\multiput(467.00,364.34)(22.000,-1.000){2}{\rule{5.300pt}{0.800pt}}
\put(511,362.84){\rule{10.600pt}{0.800pt}}
\multiput(511.00,363.34)(22.000,-1.000){2}{\rule{5.300pt}{0.800pt}}
\multiput(642.00,365.39)(4.704,0.536){5}{\rule{6.067pt}{0.129pt}}
\multiput(642.00,362.34)(31.408,6.000){2}{\rule{3.033pt}{0.800pt}}
\put(555.0,364.0){\rule[-0.400pt]{20.958pt}{0.800pt}}
\end{picture}
\caption{FB asymmetry for $\rho$=-0.07}
% GNUPLOT: LaTeX picture
\setlength{\unitlength}{0.240900pt}
\ifx\plotpoint\undefined\newsavebox{\plotpoint}\fi
\sbox{\plotpoint}{\rule[-0.200pt]{0.400pt}{0.400pt}}%
\begin{picture}(750,720)(0,0)
\font\gnuplot=cmr10 at 10pt
\gnuplot
\sbox{\plotpoint}{\rule[-0.200pt]{0.400pt}{0.400pt}}%
\put(176.0,383.0){\rule[-0.200pt]{122.859pt}{0.400pt}}
\put(176.0,68.0){\rule[-0.200pt]{0.400pt}{151.526pt}}
\put(176.0,68.0){\rule[-0.200pt]{4.818pt}{0.400pt}}
\put(154,68){\makebox(0,0)[r]{-0.3}}
\put(666.0,68.0){\rule[-0.200pt]{4.818pt}{0.400pt}}
\put(176.0,173.0){\rule[-0.200pt]{4.818pt}{0.400pt}}
\put(154,173){\makebox(0,0)[r]{-0.2}}
\put(666.0,173.0){\rule[-0.200pt]{4.818pt}{0.400pt}}
\put(176.0,278.0){\rule[-0.200pt]{4.818pt}{0.400pt}}
\put(154,278){\makebox(0,0)[r]{-0.1}}
\put(666.0,278.0){\rule[-0.200pt]{4.818pt}{0.400pt}}
\put(176.0,383.0){\rule[-0.200pt]{4.818pt}{0.400pt}}
\put(154,383){\makebox(0,0)[r]{0}}
\put(666.0,383.0){\rule[-0.200pt]{4.818pt}{0.400pt}}
\put(176.0,487.0){\rule[-0.200pt]{4.818pt}{0.400pt}}
\put(154,487){\makebox(0,0)[r]{0.1}}
\put(666.0,487.0){\rule[-0.200pt]{4.818pt}{0.400pt}}
\put(176.0,592.0){\rule[-0.200pt]{4.818pt}{0.400pt}}
\put(154,592){\makebox(0,0)[r]{0.2}}
\put(666.0,592.0){\rule[-0.200pt]{4.818pt}{0.400pt}}
\put(176.0,697.0){\rule[-0.200pt]{4.818pt}{0.400pt}}
\put(154,697){\makebox(0,0)[r]{0.3}}
\put(666.0,697.0){\rule[-0.200pt]{4.818pt}{0.400pt}}
\put(176.0,68.0){\rule[-0.200pt]{0.400pt}{4.818pt}}
\put(176,23){\makebox(0,0){0}}
\put(176.0,677.0){\rule[-0.200pt]{0.400pt}{4.818pt}}
\put(249.0,68.0){\rule[-0.200pt]{0.400pt}{4.818pt}}
\put(249,23){\makebox(0,0){0.05}}
\put(249.0,677.0){\rule[-0.200pt]{0.400pt}{4.818pt}}
\put(322.0,68.0){\rule[-0.200pt]{0.400pt}{4.818pt}}
\put(322,23){\makebox(0,0){0.1}}
\put(322.0,677.0){\rule[-0.200pt]{0.400pt}{4.818pt}}
\put(395.0,68.0){\rule[-0.200pt]{0.400pt}{4.818pt}}
\put(395,23){\makebox(0,0){0.15}}
\put(395.0,677.0){\rule[-0.200pt]{0.400pt}{4.818pt}}
\put(467.0,68.0){\rule[-0.200pt]{0.400pt}{4.818pt}}
\put(467,23){\makebox(0,0){0.2}}
\put(467.0,677.0){\rule[-0.200pt]{0.400pt}{4.818pt}}
\put(540.0,68.0){\rule[-0.200pt]{0.400pt}{4.818pt}}
\put(540,23){\makebox(0,0){0.25}}
\put(540.0,677.0){\rule[-0.200pt]{0.400pt}{4.818pt}}
\put(613.0,68.0){\rule[-0.200pt]{0.400pt}{4.818pt}}
\put(613,23){\makebox(0,0){0.3}}
\put(613.0,677.0){\rule[-0.200pt]{0.400pt}{4.818pt}}
\put(686.0,68.0){\rule[-0.200pt]{0.400pt}{4.818pt}}
\put(686,23){\makebox(0,0){0.35}}
\put(686.0,677.0){\rule[-0.200pt]{0.400pt}{4.818pt}}
\put(176.0,68.0){\rule[-0.200pt]{122.859pt}{0.400pt}}
\put(686.0,68.0){\rule[-0.200pt]{0.400pt}{151.526pt}}
\put(176.0,697.0){\rule[-0.200pt]{122.859pt}{0.400pt}}
\put(176.0,68.0){\rule[-0.200pt]{0.400pt}{151.526pt}}
\put(556,632){\makebox(0,0)[r]{For B}}
\put(578.0,632.0){\rule[-0.200pt]{15.899pt}{0.400pt}}
\put(249,256){\usebox{\plotpoint}}
\multiput(249.00,256.58)(0.579,0.498){73}{\rule{0.563pt}{0.120pt}}
\multiput(249.00,255.17)(42.831,38.000){2}{\rule{0.282pt}{0.400pt}}
\multiput(293.58,294.00)(0.498,0.569){83}{\rule{0.120pt}{0.556pt}}
\multiput(292.17,294.00)(43.000,47.846){2}{\rule{0.400pt}{0.278pt}}
\multiput(336.58,343.00)(0.498,0.556){85}{\rule{0.120pt}{0.545pt}}
\multiput(335.17,343.00)(44.000,47.868){2}{\rule{0.400pt}{0.273pt}}
\multiput(380.58,392.00)(0.498,0.522){85}{\rule{0.120pt}{0.518pt}}
\multiput(379.17,392.00)(44.000,44.924){2}{\rule{0.400pt}{0.259pt}}
\multiput(424.00,438.58)(0.499,0.498){83}{\rule{0.500pt}{0.120pt}}
\multiput(424.00,437.17)(41.962,43.000){2}{\rule{0.250pt}{0.400pt}}
\multiput(467.00,481.58)(0.549,0.498){77}{\rule{0.540pt}{0.120pt}}
\multiput(467.00,480.17)(42.879,40.000){2}{\rule{0.270pt}{0.400pt}}
\multiput(511.00,521.58)(0.594,0.498){71}{\rule{0.576pt}{0.120pt}}
\multiput(511.00,520.17)(42.805,37.000){2}{\rule{0.288pt}{0.400pt}}
\multiput(555.00,558.58)(0.629,0.498){67}{\rule{0.603pt}{0.120pt}}
\multiput(555.00,557.17)(42.749,35.000){2}{\rule{0.301pt}{0.400pt}}
\multiput(599.00,593.58)(0.673,0.497){61}{\rule{0.637pt}{0.120pt}}
\multiput(599.00,592.17)(41.677,32.000){2}{\rule{0.319pt}{0.400pt}}
\multiput(642.00,625.58)(0.688,0.497){61}{\rule{0.650pt}{0.120pt}}
\multiput(642.00,624.17)(42.651,32.000){2}{\rule{0.325pt}{0.400pt}}
\put(556,587){\makebox(0,0)[r]{For $\bar{B}$}}
\multiput(578,587)(20.756,0.000){4}{\usebox{\plotpoint}}
\put(644,587){\usebox{\plotpoint}}
\put(249,500){\usebox{\plotpoint}}
\multiput(249,500)(15.358,-13.962){3}{\usebox{\plotpoint}}
\multiput(293,460)(13.690,-15.600){4}{\usebox{\plotpoint}}
\multiput(336,411)(13.867,-15.443){3}{\usebox{\plotpoint}}
\multiput(380,362)(14.347,-14.999){3}{\usebox{\plotpoint}}
\multiput(424,316)(14.848,-14.503){3}{\usebox{\plotpoint}}
\multiput(467,274)(15.532,-13.767){2}{\usebox{\plotpoint}}
\multiput(511,235)(16.064,-13.143){3}{\usebox{\plotpoint}}
\multiput(555,199)(16.604,-12.453){3}{\usebox{\plotpoint}}
\multiput(599,166)(17.022,-11.876){2}{\usebox{\plotpoint}}
\multiput(642,136)(17.690,-10.855){3}{\usebox{\plotpoint}}
\put(686,109){\usebox{\plotpoint}}
\sbox{\plotpoint}{\rule[-0.400pt]{0.800pt}{0.800pt}}%
\put(556,542){\makebox(0,0)[r]{For B+$\bar{B}$}}
\put(578.0,542.0){\rule[-0.400pt]{15.899pt}{0.800pt}}
\put(249,380){\usebox{\plotpoint}}
\put(249,377.84){\rule{10.600pt}{0.800pt}}
\multiput(249.00,378.34)(22.000,-1.000){2}{\rule{5.300pt}{0.800pt}}
\put(293,376.84){\rule{10.359pt}{0.800pt}}
\multiput(293.00,377.34)(21.500,-1.000){2}{\rule{5.179pt}{0.800pt}}
\put(336,375.34){\rule{10.600pt}{0.800pt}}
\multiput(336.00,376.34)(22.000,-2.000){2}{\rule{5.300pt}{0.800pt}}
\put(380,373.84){\rule{10.600pt}{0.800pt}}
\multiput(380.00,374.34)(22.000,-1.000){2}{\rule{5.300pt}{0.800pt}}
\put(467,372.84){\rule{10.600pt}{0.800pt}}
\multiput(467.00,373.34)(22.000,-1.000){2}{\rule{5.300pt}{0.800pt}}
\put(511,371.84){\rule{10.600pt}{0.800pt}}
\multiput(511.00,372.34)(22.000,-1.000){2}{\rule{5.300pt}{0.800pt}}
\put(424.0,375.0){\rule[-0.400pt]{10.359pt}{0.800pt}}
\put(642,372.84){\rule{10.600pt}{0.800pt}}
\multiput(642.00,371.34)(22.000,3.000){2}{\rule{5.300pt}{0.800pt}}
\put(555.0,373.0){\rule[-0.400pt]{20.958pt}{0.800pt}}
\end{picture}
\caption{FB asymmetry for $\rho$=-0.30}
% GNUPLOT: LaTeX picture
\setlength{\unitlength}{0.240900pt}
\ifx\plotpoint\undefined\newsavebox{\plotpoint}\fi
\sbox{\plotpoint}{\rule[-0.200pt]{0.400pt}{0.400pt}}%
\begin{picture}(750,720)(0,0)
\font\gnuplot=cmr10 at 10pt
\gnuplot
\sbox{\plotpoint}{\rule[-0.200pt]{0.400pt}{0.400pt}}%
\put(176.0,383.0){\rule[-0.200pt]{122.859pt}{0.400pt}}
\put(176.0,68.0){\rule[-0.200pt]{0.400pt}{151.526pt}}
\put(176.0,68.0){\rule[-0.200pt]{4.818pt}{0.400pt}}
\put(154,68){\makebox(0,0)[r]{-0.3}}
\put(666.0,68.0){\rule[-0.200pt]{4.818pt}{0.400pt}}
\put(176.0,173.0){\rule[-0.200pt]{4.818pt}{0.400pt}}
\put(154,173){\makebox(0,0)[r]{-0.2}}
\put(666.0,173.0){\rule[-0.200pt]{4.818pt}{0.400pt}}
\put(176.0,278.0){\rule[-0.200pt]{4.818pt}{0.400pt}}
\put(154,278){\makebox(0,0)[r]{-0.1}}
\put(666.0,278.0){\rule[-0.200pt]{4.818pt}{0.400pt}}
\put(176.0,383.0){\rule[-0.200pt]{4.818pt}{0.400pt}}
\put(154,383){\makebox(0,0)[r]{0}}
\put(666.0,383.0){\rule[-0.200pt]{4.818pt}{0.400pt}}
\put(176.0,487.0){\rule[-0.200pt]{4.818pt}{0.400pt}}
\put(154,487){\makebox(0,0)[r]{0.1}}
\put(666.0,487.0){\rule[-0.200pt]{4.818pt}{0.400pt}}
\put(176.0,592.0){\rule[-0.200pt]{4.818pt}{0.400pt}}
\put(154,592){\makebox(0,0)[r]{0.2}}
\put(666.0,592.0){\rule[-0.200pt]{4.818pt}{0.400pt}}
\put(176.0,697.0){\rule[-0.200pt]{4.818pt}{0.400pt}}
\put(154,697){\makebox(0,0)[r]{0.3}}
\put(666.0,697.0){\rule[-0.200pt]{4.818pt}{0.400pt}}
\put(176.0,68.0){\rule[-0.200pt]{0.400pt}{4.818pt}}
\put(176,23){\makebox(0,0){0}}
\put(176.0,677.0){\rule[-0.200pt]{0.400pt}{4.818pt}}
\put(249.0,68.0){\rule[-0.200pt]{0.400pt}{4.818pt}}
\put(249,23){\makebox(0,0){0.05}}
\put(249.0,677.0){\rule[-0.200pt]{0.400pt}{4.818pt}}
\put(322.0,68.0){\rule[-0.200pt]{0.400pt}{4.818pt}}
\put(322,23){\makebox(0,0){0.1}}
\put(322.0,677.0){\rule[-0.200pt]{0.400pt}{4.818pt}}
\put(395.0,68.0){\rule[-0.200pt]{0.400pt}{4.818pt}}
\put(395,23){\makebox(0,0){0.15}}
\put(395.0,677.0){\rule[-0.200pt]{0.400pt}{4.818pt}}
\put(467.0,68.0){\rule[-0.200pt]{0.400pt}{4.818pt}}
\put(467,23){\makebox(0,0){0.2}}
\put(467.0,677.0){\rule[-0.200pt]{0.400pt}{4.818pt}}
\put(540.0,68.0){\rule[-0.200pt]{0.400pt}{4.818pt}}
\put(540,23){\makebox(0,0){0.25}}
\put(540.0,677.0){\rule[-0.200pt]{0.400pt}{4.818pt}}
\put(613.0,68.0){\rule[-0.200pt]{0.400pt}{4.818pt}}
\put(613,23){\makebox(0,0){0.3}}
\put(613.0,677.0){\rule[-0.200pt]{0.400pt}{4.818pt}}
\put(686.0,68.0){\rule[-0.200pt]{0.400pt}{4.818pt}}
\put(686,23){\makebox(0,0){0.35}}
\put(686.0,677.0){\rule[-0.200pt]{0.400pt}{4.818pt}}
\put(176.0,68.0){\rule[-0.200pt]{122.859pt}{0.400pt}}
\put(686.0,68.0){\rule[-0.200pt]{0.400pt}{151.526pt}}
\put(176.0,697.0){\rule[-0.200pt]{122.859pt}{0.400pt}}
\put(176.0,68.0){\rule[-0.200pt]{0.400pt}{151.526pt}}
\put(556,632){\makebox(0,0)[r]{For B}}
\put(578.0,632.0){\rule[-0.200pt]{15.899pt}{0.400pt}}
\put(249,259){\usebox{\plotpoint}}
\multiput(249.00,259.58)(0.594,0.498){71}{\rule{0.576pt}{0.120pt}}
\multiput(249.00,258.17)(42.805,37.000){2}{\rule{0.288pt}{0.400pt}}
\multiput(293.58,296.00)(0.498,0.546){83}{\rule{0.120pt}{0.537pt}}
\multiput(292.17,296.00)(43.000,45.885){2}{\rule{0.400pt}{0.269pt}}
\multiput(336.58,343.00)(0.498,0.545){85}{\rule{0.120pt}{0.536pt}}
\multiput(335.17,343.00)(44.000,46.887){2}{\rule{0.400pt}{0.268pt}}
\multiput(380.58,391.00)(0.498,0.511){85}{\rule{0.120pt}{0.509pt}}
\multiput(379.17,391.00)(44.000,43.943){2}{\rule{0.400pt}{0.255pt}}
\multiput(424.00,436.58)(0.511,0.498){81}{\rule{0.510pt}{0.120pt}}
\multiput(424.00,435.17)(41.942,42.000){2}{\rule{0.255pt}{0.400pt}}
\multiput(467.00,478.58)(0.579,0.498){73}{\rule{0.563pt}{0.120pt}}
\multiput(467.00,477.17)(42.831,38.000){2}{\rule{0.282pt}{0.400pt}}
\multiput(511.00,516.58)(0.611,0.498){69}{\rule{0.589pt}{0.120pt}}
\multiput(511.00,515.17)(42.778,36.000){2}{\rule{0.294pt}{0.400pt}}
\multiput(555.00,552.58)(0.647,0.498){65}{\rule{0.618pt}{0.120pt}}
\multiput(555.00,551.17)(42.718,34.000){2}{\rule{0.309pt}{0.400pt}}
\multiput(599.00,586.58)(0.673,0.497){61}{\rule{0.637pt}{0.120pt}}
\multiput(599.00,585.17)(41.677,32.000){2}{\rule{0.319pt}{0.400pt}}
\multiput(642.00,618.58)(0.711,0.497){59}{\rule{0.668pt}{0.120pt}}
\multiput(642.00,617.17)(42.614,31.000){2}{\rule{0.334pt}{0.400pt}}
\put(556,587){\makebox(0,0)[r]{For $\bar{B}$}}
\multiput(578,587)(20.756,0.000){4}{\usebox{\plotpoint}}
\put(644,587){\usebox{\plotpoint}}
\put(249,500){\usebox{\plotpoint}}
\multiput(249,500)(15.532,-13.767){3}{\usebox{\plotpoint}}
\multiput(293,461)(14.010,-15.314){3}{\usebox{\plotpoint}}
\multiput(336,414)(14.025,-15.300){4}{\usebox{\plotpoint}}
\multiput(380,366)(14.511,-14.840){3}{\usebox{\plotpoint}}
\multiput(424,321)(15.022,-14.323){2}{\usebox{\plotpoint}}
\multiput(467,280)(15.708,-13.566){3}{\usebox{\plotpoint}}
\multiput(511,242)(16.243,-12.921){3}{\usebox{\plotpoint}}
\multiput(555,207)(16.786,-12.208){3}{\usebox{\plotpoint}}
\multiput(599,175)(16.836,-12.138){2}{\usebox{\plotpoint}}
\multiput(642,144)(17.511,-11.143){3}{\usebox{\plotpoint}}
\put(686,116){\usebox{\plotpoint}}
\sbox{\plotpoint}{\rule[-0.400pt]{0.800pt}{0.800pt}}%
\put(556,542){\makebox(0,0)[r]{For B+$\bar{B}$}}
\put(578.0,542.0){\rule[-0.400pt]{15.899pt}{0.800pt}}
\put(249,381){\usebox{\plotpoint}}
\put(249,378.84){\rule{10.600pt}{0.800pt}}
\multiput(249.00,379.34)(22.000,-1.000){2}{\rule{5.300pt}{0.800pt}}
\put(293,377.84){\rule{10.359pt}{0.800pt}}
\multiput(293.00,378.34)(21.500,-1.000){2}{\rule{5.179pt}{0.800pt}}
\put(336,376.84){\rule{10.600pt}{0.800pt}}
\multiput(336.00,377.34)(22.000,-1.000){2}{\rule{5.300pt}{0.800pt}}
\put(380,375.84){\rule{10.600pt}{0.800pt}}
\multiput(380.00,376.34)(22.000,-1.000){2}{\rule{5.300pt}{0.800pt}}
\put(467,374.84){\rule{10.600pt}{0.800pt}}
\multiput(467.00,375.34)(22.000,-1.000){2}{\rule{5.300pt}{0.800pt}}
\put(424.0,377.0){\rule[-0.400pt]{10.359pt}{0.800pt}}
\put(599,373.84){\rule{10.359pt}{0.800pt}}
\multiput(599.00,374.34)(21.500,-1.000){2}{\rule{5.179pt}{0.800pt}}
\put(642,374.34){\rule{10.600pt}{0.800pt}}
\multiput(642.00,373.34)(22.000,2.000){2}{\rule{5.300pt}{0.800pt}}
\put(511.0,376.0){\rule[-0.400pt]{21.199pt}{0.800pt}}
\end{picture}
\end{center}
\end{figure}
for three values of the parameter $\rho$  in the experimentally 
allowed range for
     $\eta$\  =0.34.
As can be seen , there is some dependence on the value of the parameter
$\rho$.
The study of the FB asymmetry in B-decays would thus be a useful confirmatory 
data in pinning down the value of CKM matrix elements. 

The difference between the B and the $\bar{B}$ asymmetry is most pronounced,
below the J/$\psi$ threshold, as expected. This raises the possibility\
of measuring the asymmetry in a beam containing equal number of B and $\bar{B}$
particles; this would then be directly proportional to $\eta$. In figures (1)-(3),
we have shown the values expected for a mixed system. In Table -I , we show also
\begin{table}
\begin{center}
\caption{FB Asymmetry averaged between s=0.05 and s=0.35 for B, $\bar{B}$ and
         B+$\bar{B}$ \ system.}
\begin{tabular}{|p{1.5cm}|p{1.5cm}|p{1.5cm}|p{1.5cm}|}\hline
$\rho$   &   For B   &   For$\bar{B}$   &   For B+$\bar{B}$\\[.6cm] \hline\hline
0.30     &   0.086   &    -0.103        &   -0.014\\[.4cm] \hline
-0.07    &   0.086   &   -0.094         &   -0.006\\[.4cm] \hline
-0.30    &   0.080   &    -0.086        &   -0.005\\[.4cm] \hline
\end{tabular}
\end{center}
\end{table}
the value of this mixed asymmetry parameter averaged over a range of s
from 0.05
to 0.35, well below the region of the resonances.

The magnitude of the FB asymmetry is of the same of the order as the CP-asymmetry
in $B\rightarrow\ X_d\ e^+\ e^-$ and will be within observational range at future
colliders. Improvement of statistics would perhaps also the make the asymmetry
observable in a beam containing equal numbers of B and $\bar{B}$'s, which exper-
imentally would not require any 'tagging' and is thus an interesting
possibility .
\\[1. cm]
I  would like the International Centre for Theoretical Physics at Trieste
where this work was done. I would also like to thank Professor L.M.Sehgal
for going through the manuscript and for pointing out that  very
recently completed  a similar investigation whose results match the ones
quoted
here.

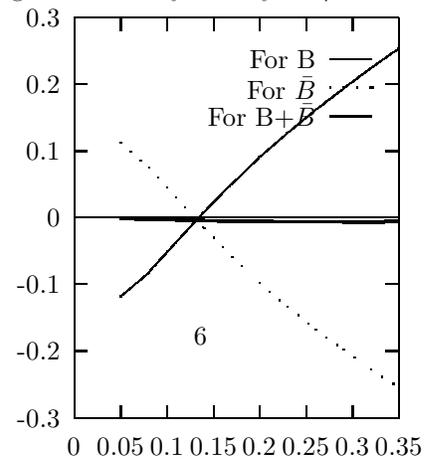
\end{document}